\documentclass[%linenumbers,
pra,reprint,superscriptaddress,nofootinbib]{revtex4-2}

\usepackage{times}% Times fonts
\usepackage{fancyhdr}
\usepackage{graphicx}% Figure package
\usepackage{float}% Float package
\usepackage{tabularx,booktabs} % Package for tables
\usepackage{enumitem}% Package for enumeration and itemization
\usepackage{multirow}
\usepackage{threeparttable}

\usepackage{physics}% Physics package including Dirac notation
\usepackage{amsmath}% Math package for miscellaneous enhancements
\usepackage{amssymb}% Package for mathbb font
\usepackage{dsfont}% Package for mathds font, e.g., \mathds{X}
\usepackage{mathrsfs} % Calligraphic math letters, e.g., \mathscr{X}
\usepackage{euscript}% Calligraphic math letters, e.g., \EuScript{X}
\usepackage{upgreek}% Package for alternative greek letters, e.g., $\Upomega$
\usepackage{mathtools}% Package for mathematical symbols, e.g., \coloneqq
\usepackage{mathdots}% Package for making dots circles
\usepackage[bbgreekl]{mathbbol}% Package for mathbbol font (use only for greek letters)
\DeclareSymbolFontAlphabet{\mathbbol}{bbold}% Declares command \mathbbol
\DeclareSymbolFontAlphabet{\mathbb}{AMSb}% Declares command \mathbb  (introduced because mathbbol package supersedes amssymb)

\usepackage{amsmath}% Math package
\usepackage{amsthm}% Theorem package
\theoremstyle{remark}% Theorem style used by APS (remark,plain,definition)
\makeatletter
\def\maketag@@@#1{\hbox{\m@th\normalfont\normalsize#1}}  
\makeatother

\usepackage[usenames,dvipsnames,table]{xcolor}% Package for new colors
\definecolor{blue_ref}{RGB}{46,48,146}% Definition of a new color for references
\usepackage{colortbl}

\usepackage[colorlinks=true,urlcolor=blue_ref,citecolor=blue_ref,linkcolor=blue_ref]{hyperref}% Hypertext package

\usepackage{comment} % Comments
\usepackage{soul} % Strike out text using \st{...}
\setstcolor{red}% Strike out color
\usepackage{cancel}% Strike out math text using \cancel{…}
\usepackage{lipsum}% Dummy text package

\usepackage{pgffor}

\usepackage{pdfpages} % Package for attaching pdfs
\makeatletter
\AtBeginDocument{\let\LS@rot\@undefined}
\makeatother

\usepackage[us,hhmmss]{datetime} % Date format package

\newcommand{\m}[1]{\mathit{#1}}% Matrix/Operator
\newcommand{\f}[1]{\mathrm{#1}}% Function
\newcommand{\map}[1]{\ifcat\noexpand#1\relax#1\else{\mathcal{#1}}\fi}% Superoperator
\newcommand{\set}[1]{\mathbbol{#1}}% Set
\newcommand{\onorm}[1]{{\left\vert\kern-0.25ex\left\vert\kern-0.25ex\left\vert #1 
    \right\vert\kern-0.25ex\right\vert\kern-0.25ex\right\vert}}
    
\newcommand{\eq}[2]{\begin{equation} \label{eq:#1} #2 \end{equation}}% Shortcut for equation
\newcommand{\al}[2]{\begin{align} \label{eq:#1} #2 \end{align}}% Shortcut for align
\newcommand{\alsub}[2]{\begin{subequations}\label{eq:#1}\begin{align} #2 \end{align}\end{subequations}}% Shortcut for align sub-equations
\let\geq\geqslant% Replace \geq with \geqslant

\newcommand{\qs}{S}% Generic quantum state
\def\TUC{School of Electrical and Computer Engineering, Technical University of Crete, Chania 73100, Greece}

\def\NUS{Centre for Quantum Technologies, National University of Singapore, 3 Science Drive 2, 117543, Singapore}

\def\NCSR{Institute for Quantum Computing and Quantum Technologies, NCSR Demokritos, Greece}

\def\Southampton{School of Electronics and Computer Science, University of Southampton, Southampton SO17 1BJ, UK}

\begin{document}

% Title
\title{Depth optimization of CNOT ladder circuits}

% Authors
\author{Spyros Tserkis}
\email{spyrostserkis@gmail.com}
\affiliation{\TUC}
\affiliation{\NCSR}

\author{Muhammad Umer}
\email{umer@u.nus.edu}
\affiliation{\NUS}

\author{Dimitris G. Angelakis}
\email{dimitris.angelakis@gmail.com}
\affiliation{\Southampton}
\affiliation{\NCSR}
\affiliation{\NUS}

% Date
%\date{\today \; at \ \currenttime}
%\date{\today}

% Abstract
\begin{abstract}
The increasing depth of quantum circuits presents a major limitation for the execution of quantum algorithms, as the limited coherence time of physical qubits leads to noise that manifests as errors during computation. In this work, we focus on CNOT ladder circuits, which find applications in several quantum computing tasks, including the preparation of GHZ states, the implementation of fan-out and long-range CNOT gates, fermionic simulations, and the construction of ansatz circuits for variational quantum computing. The linearly increasing depth of a CNOT ladder circuit can be exchanged for constant CNOT depth at the expense of wider circuits that rely on mid-circuit measurements and classically controlled operations. Our error analysis shows that the choice between these two constructions depends on the relative difference between CNOT and idling error rates. Overall, the technique developed in this work enables low-depth implementations of circuits that are ubiquitous in quantum computing algorithms.
\end{abstract}

\maketitle

\section{Introduction}
\label{sec_intro}

Implementing quantum algorithms often requires circuits with considerable depth, especially as the complexity of the computational task increases. These deep circuits typically involve highly entangled states that must retain coherence for extended durations, which presents a major challenge for current quantum hardware. Consequently, reducing circuit depth has become a central goal in quantum algorithm design, and a substantial body of research has been dedicated to this problem~\cite{Bennakhi_Byrd_Franzon_IEEE_24, Grimsley_etal_NC_19, Tang_etal_PRXQ_21, Yordanov_etal_CP_21, Sim_etal_QST_21, Hu_etal_IEEE_22, Moore_Nilsson_SIAM_01, Anders_etal_PRA_10, Proctor_Andersson_Kendon_PRA_13, Proctor_Kendon_EPJ_14, Jiang_etal_SIAM_20, Buhrman_etal_Q_24, Liu_Gheorghiu_Q_22, Watts_etal_ACM_19, Quek_Kaur_Wilde_Q_24, Baumer_etal_PRXQ_24, Baumer_Woerner_PRR_25, Cao_Eisert_PRL_26}. In particular, shallower circuits can be achieved either by optimizing the gate sequence applied to existing qubits~\cite{Bennakhi_Byrd_Franzon_IEEE_24, Grimsley_etal_NC_19, Tang_etal_PRXQ_21, Yordanov_etal_CP_21, Sim_etal_QST_21, Hu_etal_IEEE_22} or by increasing the number of qubits, which in turn often requires the use of mid-circuit measurements and classically controlled operations~\cite{Moore_Nilsson_SIAM_01, Anders_etal_PRA_10, Proctor_Andersson_Kendon_PRA_13, Proctor_Kendon_EPJ_14, Jiang_etal_SIAM_20, Buhrman_etal_Q_24, Liu_Gheorghiu_Q_22, Watts_etal_ACM_19, Quek_Kaur_Wilde_Q_24, Baumer_etal_PRXQ_24, Baumer_Woerner_PRR_25, Cao_Eisert_PRL_26}.

In this work, we adopt the latter approach for circuits consisting of controlled-NOT (CNOT) ladders, also known as CNOT staircases, i.e., sequences of consecutive CNOT gates arranged in a ladder pattern where each gate shares a qubit with the next. The core idea of our method is to substitute CNOT gates with equivalent non-unitary processes that yield the same effect on the register qubits. These non-unitary processes have the advantage of being stacked on top of each other while maintaining a shallow and constant circuit depth.

Reducing the depth of CNOT ladder circuits is a problem that has been studied previously~\cite{Watts_etal_ACM_19, Quek_Kaur_Wilde_Q_24, Baumer_Woerner_PRR_25}, so we compare our approach to existing methods. We show that the advantage of our method lies in its modularity, which allows it to handle both ascending and descending ladders. Finally, we perform an error analysis to determine whether a CNOT ladder is worth rewriting as a constant depth circuit, depending on the noise profile of the platform.

The remainder of the paper is structured as follows. In Section~\ref{sec_ladders} we define the notion of the CNOT ladder circuit and introduce the method that can decrease their depth. The usefulness of CNOT ladder circuits in quantum computing is discussed in Section~\ref{sec_apps}. In Section~\ref{sec_noise} an error analysis is presented, where the non-unitary CNOT ladder is compared to the conventional unitary circuit. Finally, the paper is concluded in Section \ref{sec_conculsion}.

\begin{table*}[tbh!]
\caption{Circuit characteristics (circuit width, CNOT depth, CNOT count, measurement count, and conditional gate count) for several constructions of a CNOT ladder circuit. Note that the expressions in row 3 correspond to an odd value $n$, while in row 4 to an even value $n$.}
\begin{ruledtabular}
\renewcommand{\arraystretch}{1.1}
\begin{tabular}{cccccccc}
Row & Type & Work & Circuit Width & CNOT depth & CNOT count & Meas. count & Cond. gate count  \\ \hline
\rule{0pt}{3ex}1 & Unitary   &  --- & $n$ & $n-1$  & $n-1$ & 0     & 0  \\
2 & Unitary   & Ref.~\cite{Remaud_Vandaele_B_25} & $n$ & $\left\lfloor \log_2 n \right\rfloor + \left\lfloor \log_2 \frac{2n}{3} \right\rfloor$  & $2n - 2 - \left\lfloor \log_2 n \right\rfloor - \left\lfloor \log_2 \frac{2n}{3} \right\rfloor$ & 0     & 0   \\
3 & Non-unitary & Ref.~\cite{Baumer_etal_PRXQ_24} & $n$\tnote{a} & 3 & $3(n-1)/2$  & $(n-1)/2$ & $(n-1)/2$ \\
4 & Non-unitary & Ref.~\cite{Quek_Kaur_Wilde_Q_24} & $n$\tnote{b} & 3 & $3n/2-2$  & $n/2-1$ & $n/2-1$ \\
5 & Non-unitary & Ref.~\cite{Baumer_Woerner_PRR_25} & $2n-1$ & 2 & $2n-2$ & $n-1$ & $n-1$  \\
6 & Non-unitary & Ref.~\cite{Quek_Kaur_Wilde_Q_24} & $2n-2$ & 2 & $2n-3$ & $n-2$ & $n-2$  \\
7 & Non-unitary & Present Work & $2n-3$ & 2  & $2n-4$ & $n-3$ & $n-2$  
\end{tabular}
\end{ruledtabular}
\label{table:comparison}
\end{table*}

\section{CNOT Ladder Circuits}
\label{sec_ladders}

\begin{figure}[b!]
\centering
\includegraphics[width=\columnwidth]{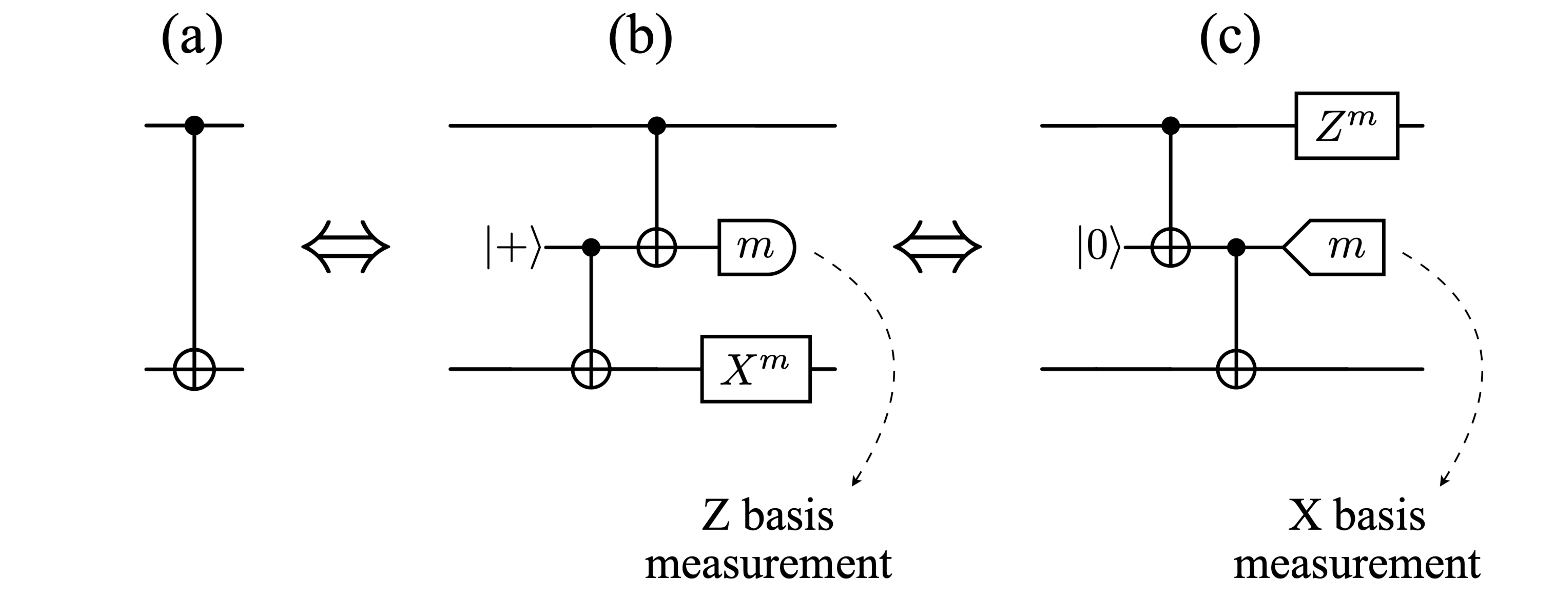}
\caption{A CNOT gate depicted in panel (a) as a unitary transformation, and in panels (b) and (c) as a non-unitary transformation accompanied with auxiliary qubits ($\ket{0}$ or $\ket{+}$), mid-circuit measurements in $\m{X}$ or $\m{Z}$ basis, and classically controlled operations.}
\label{fig:mbcnot}
\end{figure}

A CNOT gate, shown in Fig.~\ref{fig:mbcnot}(a), applies the following unitary transformation:
\eq{cnot}{
\text{CNOT} [\ket{c} \otimes \ket{t}] \quad \longmapsto \quad \ket{c} \otimes \ket{c \oplus t} \,,
}
where $\ket{c}$ corresponds to the control qubit and $\ket{t}$ to the target one. The symbol $\oplus$ indicates the XOR operation.

\begin{figure}[b!]
\centering
\includegraphics[width=\columnwidth]{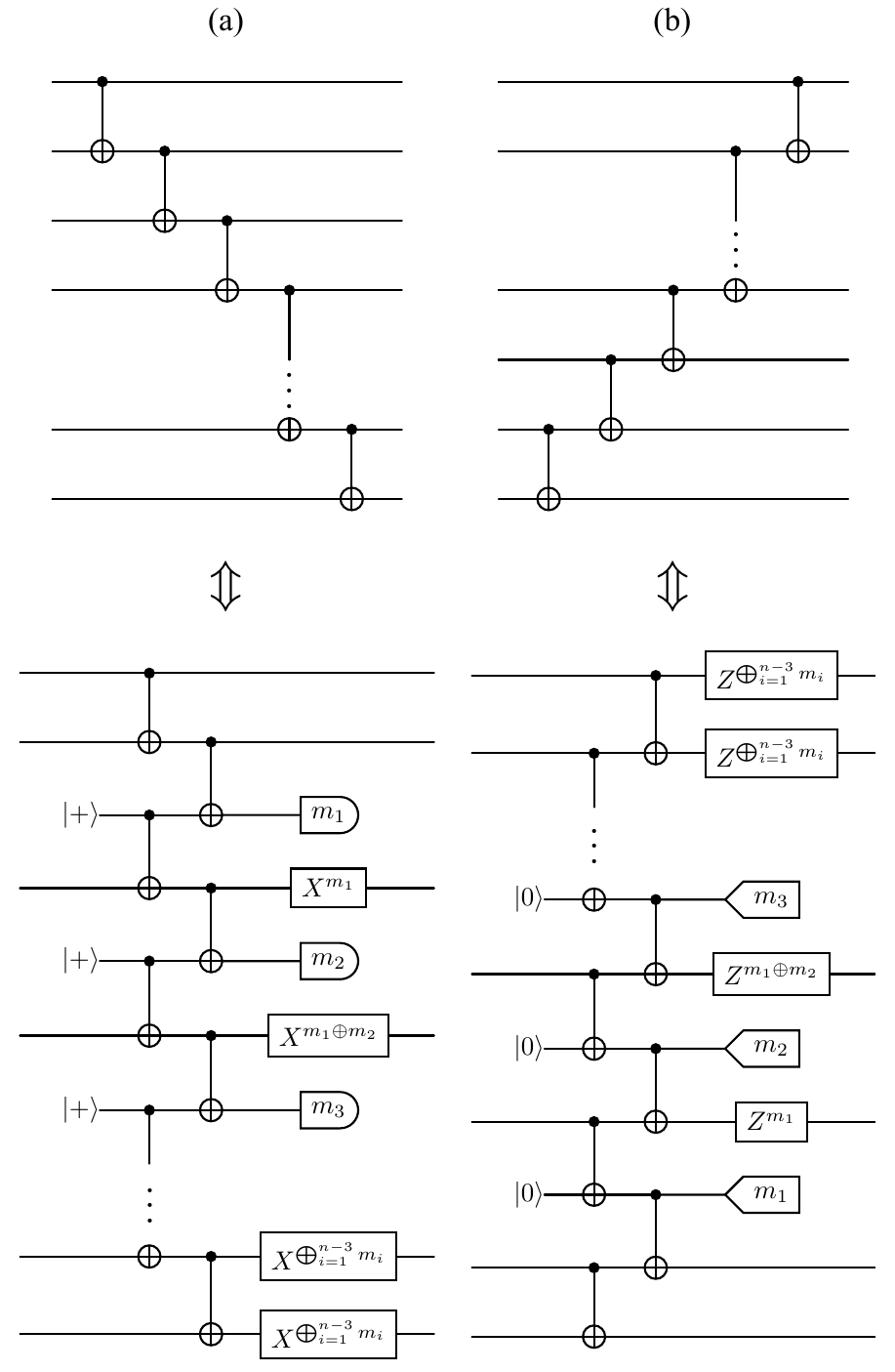}
\caption{In the top part of panel (a) a descending CNOT ladder is depicted as a unitary transformation run on a $n$-qubit circuit, while in the lower part the same ladder is depicted as a non-unitary transformation run on a $(2n{-}3)$-qubit circuit. In panel (b) the analogous circuits are presented for the ascending CNOT ladder.}
\label{fig:ladders}
\end{figure}

A CNOT ladder (or CNOT staircase) circuit consists of a sequence of consecutive CNOT gates, where each gate shares a qubit with the next one. A descending $n$-qubit CNOT ladder, shown in the upper part of Fig.~\ref{fig:ladders}(a), is defined as
\eq{}{
\text{L}_{\downarrow} \bigotimes_{i=1}^n \ket{x_i} \quad \longmapsto \quad \bigotimes_{k=1}^n \Big| \bigoplus_{j=1}^k x_j \Big\rangle \,,
}
and an ascending $n$-qubit CNOT ladder, shown in the upper part of Fig.~\ref{fig:ladders}(b), is defined as
\eq{}{
\text{L}_{\uparrow} \bigotimes_{i=1}^n \ket{x_i} \quad \longmapsto \quad \bigotimes_{k=1}^n \Big| \bigoplus_{j=k}^n x_j \Big\rangle \,.
}

\begin{figure*}[t!]
\centering
\includegraphics[width=\textwidth]{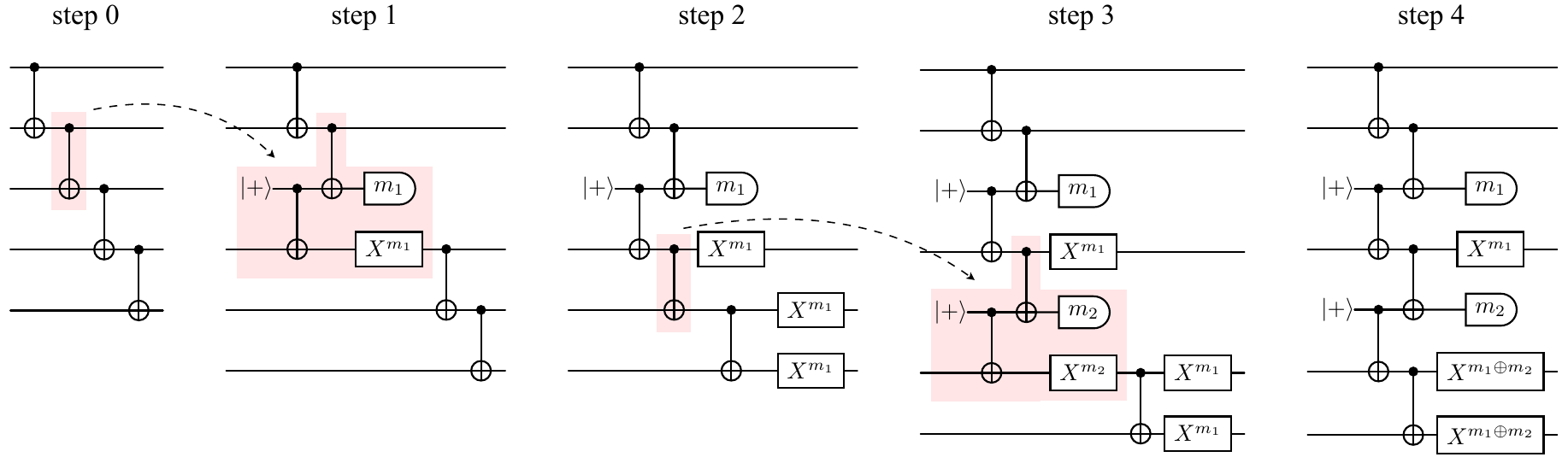}
\caption{Step by step transition from a unitary to a non-unitary circuit. Step 0 is the initial unitary circuit. Step 1 shows the first substitution of a CNOT gate with its measurement-based equivalent. In step 2 the commutation of the conditional gate takes place. In step 3 the second substitution of CNOT is shown. Finally, in step 4 the commutation of the last gate takes place.}
\label{fig:steps}
\end{figure*}

The simplest way to represent a CNOT ladder is through unitary circuits such as the ones in the upper part of Fig.~\ref{fig:ladders}. For these circuits the CNOT depth and CNOT count increase linearly, i.e., $n-1$, with the register qubit count $n$, as shown in the first row of Table~\ref{table:comparison}. Recently, an alternative construction has been proposed~\cite{Remaud_Vandaele_B_25}, that manages to represent the CNOT ladder as a unitary transformation on a circuit of equal width but with smaller CNOT depth. The circuit characteristics of that method are given in the second row of Table~\ref{table:comparison}.

Reducing the CNOT depth of a circuit and ideally keeping it constant is an active research field. Instead of re-arranging quantum gates in the circuit of the same width, there is also the option of adding auxiliary qubits or, more drastically, adding mid-circuit measurements and classically controlled operations. In rows 3-6 of Table~\ref{table:comparison} we present four such approaches found in the literature~\cite{Quek_Kaur_Wilde_Q_24, Baumer_Woerner_PRR_25, Baumer_etal_PRXQ_24}, where the CNOT ladders are implemented in non-unitary circuits.

In this work, we propose another non-unitary approach, depicted in the lower part of Fig.~\ref{fig:ladders}. The circuit characteristics of our approach are given in the last (seventh) row of Table~\ref{table:comparison}. We see that the CNOT depth can be kept constant and equal to 2 regardless of the register qubit count. The same CNOT depth is also achieved by two other methods (fifth and sixth rows in Table~\ref{table:comparison}), but in our case circuit width, CNOT count, and measurement count are smaller than existing approaches. In the following, we describe the method in detail.

The non-unitary circuits in the lower part of Fig.~\ref{fig:ladders} are constructed by substituting each CNOT gate of the unitary circuits at the top part of Fig.~\ref{fig:ladders} with its measurement-based equivalent shown in panels (b) and (c) of Fig.~\ref{fig:mbcnot}. The measurement-based CNOT gates in panels (b) and (c) of Fig.~\ref{fig:mbcnot} include an extra qubit, called auxiliary, that is initialized to a particular state, $\ket{0}$ or $\ket{+}$. The auxiliary qubit interacts with both of the register qubits introducing entanglement in the circuit, and is subsequently measured on a particular basis depending on the construction. Based on the measurement outcome, a conditional single-qubit gate $\m{X}$ or $\m{Z}$ is applied to a specific register qubit. The equivalence between the unitary and the measurement-based versions of the CNOT gate is a result that has been mentioned before in the literature~\cite{Yimsiriwattana_Lomonaco_04}, but for completeness a proof of this equivalence is provided in Appendix~\ref{appA}. Analogous constructions can easily be found for an arbitrary controlled-U (CU) gate, but the point of this work is to focus on common quantum computing gates rather than arbitrary unitary transformations\footnote{The measurement-based CNOT operation, discussed in this paragraph, is reminiscent of the gate teleportation protocol~\cite{Nielsen_Chuang_PRL_97, Gottesman_Chuang_N_99}, with the key distinction that the latter requires two auxiliary qubits which are initialized in an entangled state. In practice, starting with a CNOT gate, if we substitute the primitive non-unitary circuits twice, we retrieve the teleported CNOT gate.}.

In both CNOT ladders of Fig.~\ref{fig:ladders} the substitution begins from left to right. It is important to note that the first and last CNOT gates do not need to be taken into account in this process, as the first CNOT gate is already located at the first layer of CNOT gates that is about to be constructed, while the last CNOT gate is commuted to the second layer of CNOT gates that has been constructed at the point. In Fig.~\ref{fig:steps} we diagrammatically present how a $5$-qubit descending CNOT ladder is transformed from a unitary to a non-unitary circuit step by step. In step 0, we have the initial unitary circuit. In step 1, we substitute the second CNOT gate from left with the measurement-based CNOT gate. In step 2, we commute the rest of CNOT gates over the conditional gate that was introduced in the previous step. The commutation of CNOT gates results in two more conditional gates on the last two qubits, due to the identity $(\m{X} \otimes \m{I}) \textrm{CNOT} = \textrm{CNOT} (\m{X} \otimes \m{X})$. In step 3, we substitute the next CNOT gate with its measurement-based equivalent operation, and finally in step 4, we commute the last conditional gate.

Conceptually, this method reduces the CNOT depth of a circuit by increasing the width of the circuit and its CNOT density. The higher CNOT density allows for less idling noise, i.e., the noise induced to the circuit due to the absence of an operation while the qubits need to stay coherent. This trade-off can also be understood in terms of the active and idle volume discussed in Ref.~\cite{Litinski_Nickerson_arxiv_22}. In particular, the largest part of the circuit volume in the unitary core circuits in the upper of Fig.~\ref{fig:ladders} consists of idle volume, whereas the largest part of the circuit volume in the non-unitary core circuits in the lower of Fig.~\ref{fig:ladders} consists of active volume.

\begin{figure*}[t!]
\centering
\includegraphics[width=\textwidth]{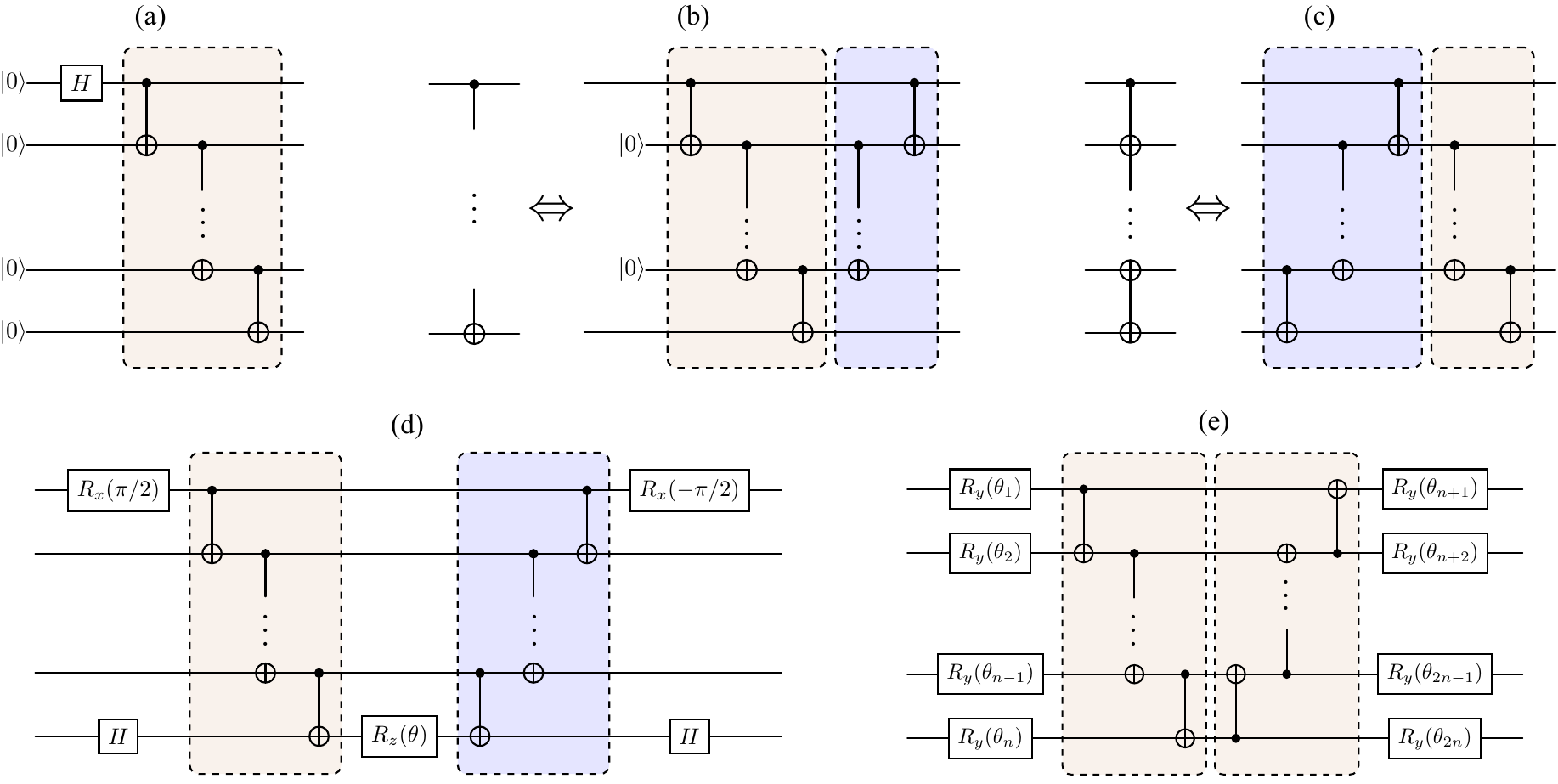}
\caption{Circuit for the construction of a (a) GHZ state, (b) long-range CNOT gate between the first and the last qubit, (c) fan-out gate, (d) spin-mapped fermionic excitation operator in fermionic simulation, and (e) ansatz circuits used in variational quantum algorithms.}
\label{fig:apps}
\end{figure*}

\section{Applications of CNOT Ladder Circuits}
\label{sec_apps}

CNOT ladder circuits appear in several applications in quantum computing. Below, we present five such examples.

\subsection{GHZ State}

A GHZ state~\cite{Greenberger_Horne_Zeilinger_B_89} is the multi-partite entangled state of the following form 
\eq{}{
\ket{\psi} = \frac{\ket{000 \cdots 0} + \ket{111 \cdots 1}}{\sqrt{2}} \,.
}
A straightforward way to create a GHZ state is, as shown in Fig.~\ref{fig:apps}(a), by initializing an array of qubits to $\ket{0}$ states, followed by a Hadamard gate on the top qubit, and a descending CNOT ladder. Low depth circuits for the construction of GHZ states have been considered before in the literature~\cite{Watts_etal_ACM_19, Liu_Gheorghiu_Q_22, Quek_Kaur_Wilde_Q_24, Baumer_etal_PRXQ_24}.

\subsection{Long-Range CNOT gate}

A long-range CNOT gate, shown in Fig.~\ref{fig:apps}(b), is a CNOT gate applied on two distant qubits, i.e.,
\eq{}{
\text{CNOT} \left[ \ket{c} \otimes \cdots \otimes \ket{t}  \right] \quad \longmapsto  \quad \ket{c} \otimes \cdots \otimes \ket{c \oplus t} \,.
}
There are several methods to achieve this task~\cite{Baumer_etal_PRXQ_24}. One of them, shown in Fig.~\ref{fig:apps}(b), requires the use of several intermediate auxiliary qubits initialized to state $\ket{0}$ and two CNOT ladders, a descending followed by one ascending one.

\subsection{Fan-Out Gate}

A fan-out gate, shown in Fig.~\ref{fig:apps}(c), is defined as
\eq{}{
\text{fan-out} \left[ \ket{c} \otimes \bigotimes_{i=1}^{n-1} \ket{t_i} \right] \quad \longmapsto  \quad \ket{c} \otimes \bigotimes_{i=1}^{n-1} \ket{c \oplus t_i} \,.
}
In Fig.~\ref{fig:apps}(c) we provide a possible construction of a fan-out gate using two CNOT ladders, found in Ref.~\cite{Baumer_Woerner_PRR_25}.

\subsection{Fermionic Simulations}

In a single-dimensional fermionic chain encoded to spin system via the Jordan–Wigner transformation, the implementation of operators such as 
\eq{}{
\exp \left[ -\imath \theta \m{X}_i \m{Y}_k \prod_{j=i+1}^{k-1} \m{Z}_j \right]
}
is mapped not only to endpoint Pauli $X_{i}$ or $Y_{k}$ operators but also to an intermediate parity string $\prod_{j=i+1}^{k-1}Z_{j}$ \cite{Nys2023}. The role of this parity string is to count the parity of all fermions lying between the two sites and thereby enforce the correct minus sign associated with fermionic anti-commutation. In quantum computation, this nonlocal parity string is commonly constructed using a CNOT ladder \cite{Zhu2018}. The sequence of nearest-neighbor CNOTs accumulates the intermediate parity onto the endpoint, a local rotation implements the desired phase, and the CNOTs are then reversed to uncompute the parity information, as illustrated in Fig. \ref{fig:apps}(d).

\subsection{Variational Quantum Algorithm Ansatz}

Variational quantum algorithms~\cite{Cerezo_etal_NRP_21} rely on parameterized quantum circuits, known as ansatz circuits, designed to prepare quantum states that approximate the solution to a given problem. Those ansatz circuits contain rows of single-qubit rotations, that are optimized according to the problem. Ansatz circuits admit multiple designs~\cite{Kandala2017, Sim_Johnson_Aspuru_AQT_19}, but a CNOT ladder structure such as the one shown in Fig.~\ref{fig:apps}(e) is commonly used~\cite{Umer_Mastorakis_Angelakis_QST_25}.

\begin{figure*}[tbh!]
\centering
\includegraphics[width=\textwidth]{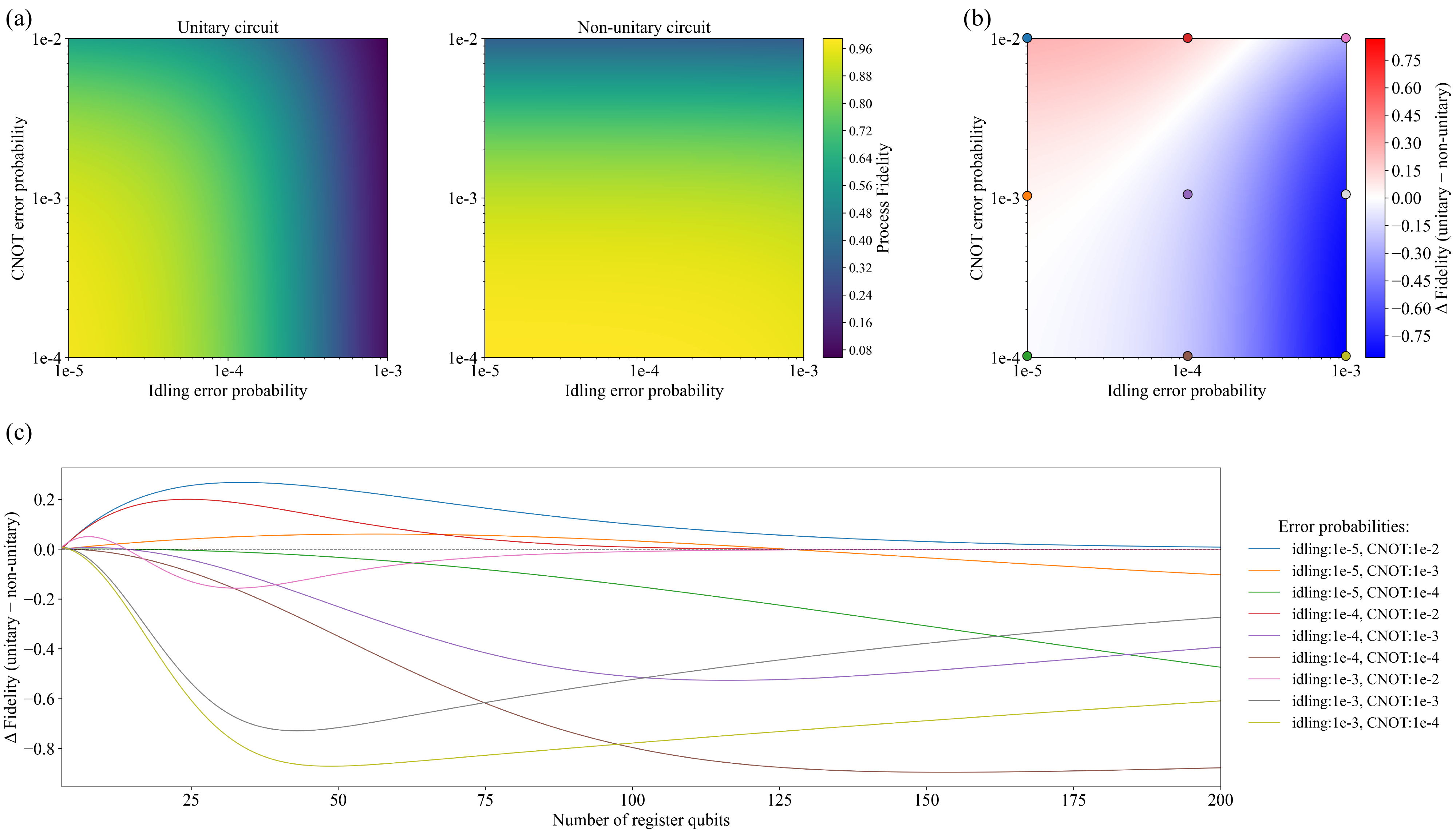}
\caption{In panel (a) two plots are presented for the unitary and the non-unitary version of the CNOT ladder. Both have the same range in terms of idling error probability and CNOT error probability: 1e-5 to 1e-3 and 1e-4 to 1e-2, respectively. The colorbar corresponds to both plots and indicates the lower bound of process fidelity of the entire circuit, which appears to be at its highest for the unitary circuit when idling error probability is low, and for the non-unitary circuit when CNOT error probability is low. In panel (b) the difference of the process fidelities is shown, $\Delta$ Fidelity, where positive (red) indicates an advantage for the unitary circuit and negative (blue) for the non-unitary one. Finally, in panel (c) we present the $\Delta$ Fidelity against the number of register qubits for nine different error probabilities, which are also indicated as colored dots on the plot of panel (b).}
\label{fig:results}
\end{figure*}

\section{Error Analysis: Results and Discussion}
\label{sec_noise}

The noise model we consider in this work is based on the one presented in Ref.~\cite{vandenBerg_etal_NP_23}. Given a quantum state with density matrix $\qs$, the noisy process is represented by a Pauli quantum channel $\map{\Uplambda} (\qs)$ as follows
\eq{}{
\map{\Uplambda}(\qs) =  (1 - p) \qs +  p P \qs P^{\dagger}   \,,
}
where $P$ belongs to the Pauli group $\set{P}$, and $p = \frac{1 - e^{-2 \lambda}}{2}$ is the error probability with $\lambda \in \set{R}^+$. This noise model assumes that a twirling approximation has been applied to remove non-Pauli noise, which is typical for error analysis in quantum computing~\cite{Bennett_etal_PRA_96, Emerson_etal_S_07, Silva_etal_PRA_08, Dankert_etal_PRA_09, Geller_Zhou_PRA_13, Wallman_Emerson_PRA_16, Cai_Benjamin_SR_19, Ware_etal_PRA_21, Hashim_etal_PRX_21} even though in real quantum computers non-Pauli noise is also expected to be present.

Given the twirling approximation, all noise elements, e.g., $\m{X}$ gates corresponding to bit-flips or $\m{Z}$ gates corresponding to phase-flips etc., can be commuted to the very end of the circuit, allowing us to more easily approximate its total impact. In order to do so, and following the analysis of Ref.~\cite{Baumer_etal_PRXQ_24}, all the errors occurred throughout the circuit can be aggregated in a total decoherence parameter $\lambda_{\text{tot}}$ as follows:
\eq{}{
\lambda_{\text{tot}} = t_{\text{idle}} \lambda_{\text{idle}} + \kappa_{\text{CNOT}} \lambda_{\text{CNOT}} + \kappa_{\text{meas}} \lambda_{\text{meas}} + \kappa_{\text{in}} \lambda_{\text{in}} + \kappa_{\text{con}} \lambda_{\text{con}} \,.
}
In the above expression, $t_{\text{idle}}$ corresponds to the time-steps where coherence needs to be maintained for quantum states while no other operation is implemented. The decoherence per idle time-step is associated to the parameter $\lambda_{\text{idle}}$. The total number of CNOT gates is given by $\kappa_{\text{CNOT}}$ and the corresponding decoherence with $\lambda_{\text{CNOT}}$. Analogously we have $\kappa_{\text{meas}}$ and $\lambda_{\text{meas}}$ for the mid-measurements, $\kappa_{\text{in}}$ and $\lambda_{\text{in}}$ for qubit initialization, and $\kappa_{\text{con}}$ and $\lambda_{\text{con}}$ for conditional gates. It is worth noting that the parameter $\lambda$ for each element in our model is versatile with respect to the underlying hardware platform because it assumes the worst case scenario that all possible Pauli errors of the corresponding group are applied with a given error probability. 

Using the total decoherence parameter, $\lambda_{\text{tot}}$, we can estimate a lower bound for the process fidelity of the entire circuit~\cite{Baumer_etal_PRXQ_24}, using the following inequality
\eq{lower_bound}{
\f{F}_{\text{pro}} \geq e^{- \lambda_{\text{tot}}} \,.
}
This benchmark choice comes as a compromise, since for circuits with very large width and depth it is practically impossible to simulate the effect of quantum noise on classical computers and calculate the exact value of fidelity. For example, in Fig.~\ref{fig:results}(c) we consider circuits with up to 200 register qubits, and thus in total around 400 qubits including the auxiliary ones, accompanied with multiple measurements and classically conditioned gates. However, the bound above allows us to compare on an equal footing the impact of decoherence and draw useful conclusions.

In Fig.~\ref{fig:results}, we compare the conventional CNOT ladder which is represented as a unitary transformation, shown in the upper part of Fig.~\ref{fig:ladders}, with the corresponding non-unitary transformation in the lower part of Fig.~\ref{fig:ladders}. In particular, in Fig.~\ref{fig:results}, we consider the case of 50 register qubits and present in panel (a) two plots whose color indicates the lower bound of the process fidelity of the entire circuit, given in the right-hand side of expression \eqref{eq:lower_bound}. Note that the color bar corresponds to both of them for an easier comparison. On the x-axis we have the idling error probability $p_{\text{idle}}$ ranging from 1e-5 to 1e-3, while on the y-axis the CNOT error probability $p_{\text{CNOT}}$ ranging from 1e-4 to 1e-2. We assume that each conditional single-qubit gate has an error probability equal to $p_{\text{con}} = p_{\text{CNOT}}/10$. Following the same reasoning, we assume that the qubit initialization and measurement error probabilities are also one order of magnitude smaller than $p_{\text{CNOT}}$, i.e., $p_{\text{in}} = p_{\text{CNOT}}/10$ and $p_{\text{meas}} = p_{\text{CNOT}}/10$. For conditional gates we assume that their decoherence is given by $\lambda_{\text{con}}=(\lambda_{\text{idle}} + \lambda_{\text{X}})/2$, as they are not applied all the time. Finally, $t_{\text{idle}}$ for the non-unitary ladder is equal to 4, while for the unitary one is equal to $n^2-3n+2$.

We observe that the unitary circuit performs the best for low idling error, while the opposite is true for the non-unitary circuit. In order to compare them quantitatively, we present in Fig.~\ref{fig:results}(b) another plot but now the colorbar corresponds to the $\Delta$ Fidelity, i.e., the process fidelity of the unitary circuit minus the non-unitary one. Positive values, represented with red color, imply a better performance for the unitary circuit, and  negative values, represented with blue color, imply a better performance for the non-unitary ones.

In Fig.~\ref{fig:results}(c) we select nine pairs for idling and CNOT error probabilities and plot the $\Delta$ Fidelity against register qubits in the range up to 200. In practice, this plot shows how Fig.~\ref{fig:results}(b) evolves for different numbers of register qubits. We observe that the non-unitary construction is preferable for most of the error parameters considered in this plot, however the corresponding trends are not the same. This implies that the choice between a unitary and a non-unitary circuit depends on the number of register qubits and the error characteristics of the platform.

\section{Conclusion}
\label{sec_conculsion}

In summary, we have presented a method for constructing shallow depth CNOT ladder circuits by replacing standard CNOT gates with their measurement-based equivalent ones, corresponding to three-qubit circuit primitives that use auxiliary qubits, mid-circuit measurements, and conditional single-qubit gates. This approach addresses the depth limitations imposed by finite coherence times in current quantum hardware and is most effective for platforms where CNOT error probabilities are relatively low compared to qubit idling error probabilities. The types of circuits considered in this work are immediately relevant to various applications, such as the preparation of GHZ states, the implementation of fan-out and long-range CNOT gates, fermionic simulations, and the construction of ansatz circuits for variational quantum computing. Overall, the technique discussed in this work enables a scalable implementation of structured circuits within the constraints of near-term quantum devices.

\section{Acknowledgements} 

This work is supported by the National Research Foundation, Singapore, and A*STAR under its CQT Bridging Grant and by the EU HORIZON—Project 101080085 – QCFD

\appendix
\section{Measurement-Based Gates}
\label{appA}

In this section we prove that the circuit in Fig.~\ref{fig:mbcnot}(a) corresponds to a CNOT operation between the first and third qubits shown in Fig.~\ref{fig:mbcnot}(b). 

We begin by considering two arbitrary states for qubit 1 and 3, i.e., $\sum_{ij} \alpha_{ij} \ket{i}_1 \ket{j}_3$, and a fixed state for qubit 2, i.e., $\ket{+}_2 = (\ket{0}_2 + \ket{1}_2)/\sqrt{2}$, so the combined input state is written as follows:
\eq{}{
\ket{\psi_0} = \frac{1}{\sqrt{2}} \sum_{ij} \alpha_{ij} \ket{i}_1 \left( \ket{0}_2 + \ket{1}_2 \right) \ket{j}_3 \,.
}
We, then, apply a CNOT gate, using qubit 2 as control and qubit 3 as target. The resulting state is
\eq{}{
\ket{\psi_1} = \frac{1}{\sqrt{2}} \sum_{ij} \alpha_{ij} \left( \ket{i}_1 \ket{0}_2 \ket{j}_3  + \ket{i}_1 \ket{1}_2 \ket{j \oplus 1}_3 \right) \,.
}
Next, we apply a CNOT gate, using qubit 1 as control and qubit 2 as target: 
\eq{}{
\ket{\psi_2} = \frac{1}{\sqrt{2}} \sum_{ij} \alpha_{ij} \left( \ket{i}_1 \ket{i}_2 \ket{j}_3  + \ket{i}_1 \ket{i \oplus 1}_2 \ket{j \oplus 1}_3 \right) \,.
}

The next operation is a measurement on the second qubit, which corresponds to a projection onto the computational basis $ \ket{m} \in \{\ket{0}, \ket{1} \}$: 
\al{}{
\ket{\psi_3} = \frac{1}{\sqrt{2}} \sum_{ij} \alpha_{ij} \big( &\ket{i}_1 \ketbra{m}{m} i \rangle_2 \ket{j}_3 \nonumber \\
&+  \ket{i}_1 \ketbra{m}{m} i \oplus 1\rangle_2 \ket{j \oplus 1}_3 \big)  \,.
}
For $m = 0$ and $m = 1$ we have respectively: 
\alsub{}{
\ket{\psi_3} &= \frac{1}{\sqrt{2}} \sum_{j} \big( \alpha_{0j}  \ket{0}_1 \ket{0}_2 \ket{j}_3 + a_{1j} \ket{1}_1 \ket{0}_2 \ket{j \oplus 1}_3 \big) \,, \\
\ket{\psi_3} &= \frac{1}{\sqrt{2}} \sum_{j} \big( \alpha_{1j}  \ket{1}_1 \ket{1}_2 \ket{j}_3 + a_{0j} \ket{0}_1 \ket{1}_2 \ket{j \oplus 1}_3 \big) \,,
} 
which can be written more compactly as
\al{}{
\ket{\psi_3} = \frac{1}{\sqrt{2}} \sum_{j} \big( &\alpha_{0j}  \ket{0}_1 \ket{m}_2 \ket{j \oplus m}_3  \nonumber \\
&+ a_{1j} \ket{1}_1 \ket{m}_2 \ket{j \oplus m \oplus 1}_3 \big) \,.
}

When the projection is equal to $\ket{m}_1 = \ket{0}_1$, we do not perform any other operation on the circuit, i.e.,
\eq{}{
\ket{\psi_4} = \frac{1}{\sqrt{2}} \sum_{j} \big( \alpha_{0j}  \ket{0}_1 \ket{0}_2 \ket{j}_3 + a_{1j} \ket{1}_1 \ket{0}_2 \ket{j \oplus 1}_3 \big) \,,
} 
but when $\ket{m}_1 = \ket{1}_1$, we apply an $\m{X}$ gate to the third qubit, i.e., 
\eq{}{
\ket{\psi_4} = \frac{1}{\sqrt{2}} \sum_{j} \big( \alpha_{0j}  \ket{0}_1 \ket{1}_2 \ket{j}_3 + a_{1j} \ket{1}_1 \ket{1}_2 \ket{j \oplus 1}_3 \big) \,.
} 
Tracing out the second qubit, we arrive at 
\al{}{
\ket{\psi_5} &= \sum_{j} \big( \alpha_{0j}  \ket{0}_1 \ket{j}_3 + a_{1j} \ket{1}_1 \ket{j \oplus 1}_3 \big) \nonumber \\
&= \sum_{kj} \alpha_{kj}  \ket{k}_1 \ket{j \oplus k}_3  \,,
}
which is the CNOT transformation on qubits 1 and 3, thus completing the proof.

The proof for the circuit in Fig.~\ref{fig:mbcnot}(c) is analogous, so we omit it for brevity.

\bibliography{Bibliography/bibliography}

\end{document}